\documentclass{emulateapj}

\newcommand\bmath[1] {\mbox{\boldmath$\rm #1$}}

\newcommand\eg{{e.g.}} 
\newcommand\ie{{i.e.}} 

\newcommand\yr{{\rm yr}} 
\newcommand\Myr{{\rm M}\yr} 

\renewcommand\d{{\rm d}}
\newcommand\e{{\rm e}}

\begin{document}

\shorttitle{Embedded Oscillating Starless Cores}
\shortauthors{Broderick et al.}

\title{The Damping Rates of Embedded Oscillating Starless Cores}

\author{Avery E. Broderick\altaffilmark{1}, Ramesh Narayan, Eric Keto \& Charles J. Lada}
\affil{Harvard-Smithsonian Center for Astrophysics, 60 Garden
  Street, Cambridge, MA 02138, USA; aeb@cita.utoronto.ca,
  rnarayan@cfa.harvard.edu, keto@cfa.harvard.edu, clada@cfa.harvard.edu}
\altaffiltext{1}{Current address: Canadian Institute for Theoretical Astrophysics, 60 St.~George
  St., Toronto, ON M5S 3H8, Canada}

\begin{abstract}
In a previous paper we demonstrated that non-radial hydrodynamic
oscillations of a thermally-supported (Bonnor-Ebert) sphere embedded
in a low-density, high-temperature medium persist for many periods.
 The predicted column density variations and molecular
spectral line profiles are similar to those observed in the Bok
globule B68 suggesting that the motions in some starless cores may
be oscillating perturbations on a thermally supported equilibrium
structure.  Such oscillations can produce molecular line maps which
mimic rotation, collapse or expansion, and thus could make determining
the dynamical state from such observations alone difficult.

However, while B68 is embedded in a very hot, low-density medium, many starless
cores are not, having interior/exterior density contrasts closer to
unity.  In this paper we investigate the oscillation damping rate as a
function of the exterior density.  For concreteness we use the same
interior model employed in Broderick et al. (2007), with varying
models for the exterior gas.  We also develop a simple analytical
formalism, based upon the linear perturbation analysis of the
oscillations, which predicts the contribution to the damping rates due
to the excitation of sound waves in the external medium.  We find that
the damping rate of oscillations on globules in dense molecular
environments is always many periods, corresponding to hundreds of
thousands of years, and persisting over the inferred lifetimes of the
globules.
\end{abstract}

\keywords{stars: formation, hydrodynamics, ISM: globules, ISM: clouds}

\section{Introduction}
The small dark molecular clouds known as starless cores are 
are significant in the interstellar medium as the potential birthplaces
of stars \citep[review by ][]{BerginTafalla2007}. As their name
implies, the starless cores do not yet contain stars, but their
properties are nearly the same as similar small clouds that do
\citep{MyersLinkeBenson1983, MyersBenson1983, BensonMyers1989}. 
Furthermore, there is a compelling similarity in the mass
function of the starless cores \citep{Lada2008}
and the initial mass function (IMF) of stars.

Observations of Bok globules (Bok 1948) and starless cores
\citep{Ward-Thompson1994,Tafalla1998,LeeMyers1999,LeeMyersTafalla2001,AlvesLadaLada2001,Ward-ThompsonMotteAndre1999,Bacmann2000,Shirley2000,Evans2001,ShirleyEvansRawlings2002,Young2003,Tafalla2004,Keto2004} 
suggest that many of these small (M $< 10$ M$_\odot$) clouds are well
described as quasi-equilibrium structures supported mostly by thermal
pressure, approximately Bonnor-Ebert (BE) spheres
\citep{Bonnor1956}. Observed molecular spectral lines
\citep{Zhou1994,Wang1995,Gregersen1997,Launhardt1998,GregersenEvans2000,LeeMyersTafalla1999,Williams1999,Lada2003,LeeMyersPlume2004,LeeBerginEvans2004,Keto2004,Crapsi2004,Sohn2004,Aguti2007}
show complex profiles that further suggest velocity and density
perturbations within these cores. We have previously shown that in at
least one case, B68, these profiles can be produced by the non-radial
oscillations of an isothermal sphere \citep{Keto2006, Broderick2007}.

To match the spectral line profiles, we have previously found that the
oscillations have to be large enough (25 \%) that the amplitudes are
non-linear \citep{Keto2006}. If such large amplitude oscillations are to be a
viable explanation for the observed complex velocity patterns then
their decay rate must be no faster than the sound-crossing or
free-fall times of the globules, $10^{5-6}$ yrs.  Numerical
simulations in which the external medium is substantially less dense
than the gas inside the core have found that mode damping is
sufficiently slow that modes will persist for many periods
\citep{Broderick2007}.

However, it is not clear that this description is appropriate for many
starless cores.  B68 is one of only a handful of cores surrounded by
hot, rarefied gas in the Pipe Nebula, where the vast majority of cores,
like those in the Taurus \& Perseus molecular clouds, appear to be
embedded in cold, dense molecular gas \citep{Lada2008,Goldsmith2008}.  Embedded
oscillating clouds will generally act as sonic transducers, exciting
sound waves in the external medium and thereby loosing energy.  A
simple one-dimensional analysis, discussed in detail in the
Appendices, suggests that when the density contrast between the core
interior and its bounding medium is close to unity this process can be
very efficient, potentially limiting the lifetime of oscillations.

In this paper we investigate the damping rate of oscillations of
isothermal spheres as a function of the density contrast with the
external bounding medium. We do this using numerical simulations of a
large amplitude oscillation superimposed upon an isothermal sphere.
To evaluate our results we also derive analytical and semi-analytical
estimates of the damping rate associated with the excitation of sound
waves in the exterior medium.  A brief discussion of the numerical
simulation is discussed in section \ref{sec:NHM}, the presentation and
discussion of the numerical results are in section \ref{sec:RaD}, and
conclusions can be found in section \ref{sec:C}.  The details of the
analytical and semi-analytical computation of the damping rates of
small amplitude oscillations are relegated to the appendices.

We find that a reduced density contrast between the core and the
exterior bounding medium does result in increased dissipation of
oscillations as momentum and energy are transferred out of the core
through the boundary.  However, if the oscillations are non-radial the
dissipation through the boundary is always less than the dissipation
rate due to mode-mode coupling.

\section{Numerical Hydrodynamic Model} \label{sec:NHM}
\begin{figure}[t!]
\centerline{
\includegraphics[width=\columnwidth]{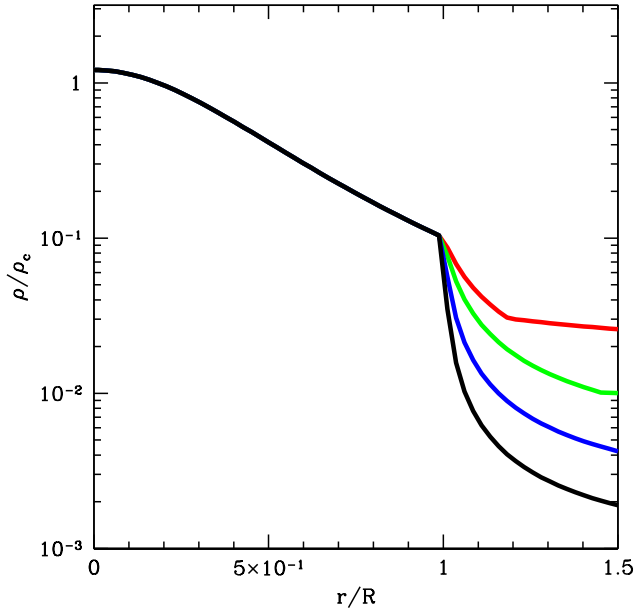}
}
\caption{The initial, unperturbed density profile of the embedded
  isothermal sphere.  The
  density contrast for each is roughly 50 (black), 25 (blue), 10
  (green) and 4 (red).  This color scheme will be consistent
  throughout the other figures.} \label{fig:ds}
\end{figure}

\begin{deluxetable}{cccc}
\tabletypesize{\scriptsize}
\tablecaption{External Gas Properties and Damping Rates \label{tab:T}}
\tablewidth{0pt}
\tablehead{
\colhead{Model} &
\colhead{$T_{\rm ext}/T$} &
\colhead{$\rho_{\rm ext}/\rho_{\rm surf}$\tablenotemark{a}} &
\colhead{$\tau_A$ [$(GM/R^3)^{-1/2}$\tablenotemark{b}]}
}
\startdata
black & $10^6$ & 0.017 & 8.5\\
blue  & $300$ & 0.037 & 6.5\\
green & $10$ & 0.087 & 4.8\\
red & $3$ & 0.22 & 3.44\\
\enddata
\tablenotetext{a}{As measured on the computational domain}
\tablenotetext{b}{$(GM/R^3)^{-1/2} \simeq 1.24\times10^5\,\yr$}
\end{deluxetable}

Because the thermal gas heating and cooling time, via collisional coupling
to dust at high densities \citep{Burke1983} and molecular line
radiation at low densities \citep{Goldsmith2001}, is short
in comparison to the typical oscillation period
(on the order of $10^5\,\yr$), the oscillations of dark-starless cores are well modeled
with an isothermal equation of state \citep{KetoField2005}.
In our simulations, we use a barotropic equation of state with an adiabatic
index of unity (as opposed to $5/3$, for example), in which case the
isothermal evolution is also adiabatic.  This allows us to replace the
energy equation with a considerably simpler adiabatic constraint.
Since the unbounded isothermal gas sphere is unstable, we truncate the
solution at a given center-to-edge density contrast ($\rho(R)/\rho_c\simeq12$,
not to be confused with the interior/exterior density contrast),
producing a stable Bonnor-Ebert sphere solution with radius ($R$) and
mass determined by the central density ($\rho_c$) and temperature ($T$).
This is done by inducing a phase change in the equation of state, \ie,
\begin{equation}
P =
\left\{
\begin{array}{rl}
\rho \frac{kT}{\mu m_p} & \mbox{if } \rho\ge\rho(R)\,,\\
\left[ \frac{\rho}{\rho(R)} \right]^\epsilon \rho(R) \frac{k T}{\mu m_p} & \mbox{if } \rho(R) > \rho \ge \beta\rho(R)\frac{T}{T_{\rm ext}}\,,\\
\rho \frac{kT_{\rm ext}}{\mu m_p} & \mbox{otherwise}\,,
\end{array}
\right.
\end{equation}
where the intermediate state, with $\beta=0.8$ and $\epsilon =
\log(\beta)/\log(\beta T/T_{\rm ext}) \ll 1$ chosen to make $P(\rho)$
continuous, is introduced to avoid numerical artifacts at the
surface, and $T_{\rm ext}\gg T$.  The specific
values of $T_{\rm ext}$ that we chose, their corresponding density
contrasts and decay timescales, are presented in Table \ref{tab:T}.  The
associated radial density profiles are shown in Figure \ref{fig:ds}.  Note that
there is a transition region which distributes the sudden drop in density
over many grid zones, though the pressure is continuous (necessarily)
and nearly constant outside the surface of the Bonnor-Ebert sphere.

We employ the same three-dimensional,
self-gravitating hydrodynamics code as in \citet{Broderick2007} to
follow the long term mode evolution.  Details regarding the numerical
algorithm and specific validation for oscillating gas spheres (in that
case a white dwarf) can be found in \citet{BroderickRathore2006}, and
thus will only be briefly summarized here.

The code is a second-order accurate (in space and time) Eulerian
finite-difference code, and has been demonstrated to have a low
diffusivity.  Because the equation of state is barotropic, we use the
gradient of the enthalpy instead of the pressure in the Euler equation
since this provides better stability in the unperturbed configuration
\citep{BroderickRathore2006}.  The Poisson equation is solved via
spectral methods, with boundary conditions set by a multipole
expansion of the matter on the computational domain.
As described in \citet{Broderick2007}, the code was validated for
this particular problem finding convergence by a resolution of $128^3$
grid zones.

The initial perturbed states are computed in the linear approximation
using the standard formalism of adiabatic \citep[which in this case is also
isothermal, ][]{Keto2006} stellar oscillations \citep{Cox1980}.
The initial conditions of the perturbed gas sphere are then given by
the sum of the equilibrium state and the linear perturbation.  We
employ the same definition for the dimensionless mode amplitude as
given in the Appendix of \citet{Keto2006}.

Throughout the evolution, the amplitude of a given oscillation mode,
denoted by its radial ($n$) and angular ($l$ \& $m$) quantum numbers,
may be estimated by the explicit integral:
\begin{equation}
A_{nlm} = \omega_{nlm}^{-1}\int \d^3\!x \,\rho\, \bmath{v}\cdot\bmath{\xi}_{nlm}^\dagger\,,
\label{eq:amp}
\end{equation}
where $\bmath{v}$ is the gas velocity, $\bmath{\xi}_{nlm}$ is the mode
displacement eigenfunction, $\omega_{nlm}$ is the mode frequency and
the dagger denotes Hermitian conjugation.\footnote{For stellar
  oscillations the mode amplitudes may be determined from the density
  perturbation alone using the velocity potential.  However, when
  there is a non-vanishing surface density, as is the case for the
  Bonnor-Ebert sphere, this is not possible and the velocity field
  must be used.}

\section{Results \& Discussion} \label{sec:RaD}
In principle, a number of hydrodynamic mechanisms exist by which the
oscillations can be damped.  In the absence of a coupling to an external
medium, these are dominated by nonlinear mode--mode coupling, in which
large scale motions excite smaller scale perturbations
\citep{Broderick2007}.  However, when embedded in a dense external
medium it is possible for the mode to damp by exciting motions in the
exterior gas.  That is, it is possible for the oscillating sphere to act
as a transducer, generating sound waves in the external gas which then
propagate outward resulting in a net radial energy flux and damping
the oscillations.

Generally, this will be a strong function of the density contrast
between the interior of the core and the exterior medium.  If the
external density has a low density, and thus little inertia, outwardly
moving sound waves will contain little energy density despite their
large amplitude and thus inefficiently damp the mode (as is the case
for B68).   Similarly, if the exterior medium has a very high density
the excited outgoing waves will have small amplitudes, which despite
the large gas inertia will again carry away only a small amount of
energy.  Conversely, when the density contrast is nearly unity, there
will be an efficient coupling between the interior and exterior waves,
resulting in a rapid flow of energy out of the cloud.

This may be made explicit via the standard three-wave analysis, in which
propagating wave solutions for the incident, reflected and transmitted
waves on either side of the density continuity are inserted into the
continuity and Euler equations, which can subsequently be solved for the
reflected and transmitted energy flux.  In this idealization the ratio
of the reflected to transmitted flux is $4\zeta/(1+\zeta)^2$, where
\begin{equation}
\zeta\equiv\rho_e c_{s,e} / \rho_i c_{s,i} = \sqrt{\rho_e/\rho_i}\,,
\end{equation}
is the ratio of the sonic impedances on either side of the
discontinuity \citep{LandauLifshitz1987}.  As anticipated, when the density of the external gas is
comparable to the surface density of the isothermal sphere we may
expect efficient transmission of sound waves from the interior to the
exterior (\ie, conversion of the oscillation into traveling waves in
the exterior), and therefore efficient damping of the pulsations.

This conclusion is also supported by a 1D analysis of the decay of a
standing wave confined to a high density region (section
\ref{sec:1DT}), for which the damping rate, $\gamma$ is given by
\begin{equation}
\gamma
=
\frac{c_{s,i}}{2 R}\ln\left(\frac{|1-\zeta|}{1+\zeta}\right)
\simeq
\frac{c_{s,i}}{R} \zeta
\,,
\end{equation}
where $c_{s,i}$ is the sound speed in the interior and $R$ is the
radius of the high-density region.  Thus, as expected lower external
densities (corresponding to smaller $\zeta$) have lower damping rates
and larger damping timescales.  Note that when $\zeta$ is of order
unity the wave decays on the sound crossing time of the dense region,
as would be expected if, e.g., we were to decompose the standing wave
into traveling waves.

\begin{figure}[t!]
\vspace*{-15mm}
\centerline{
\includegraphics[width=\columnwidth]{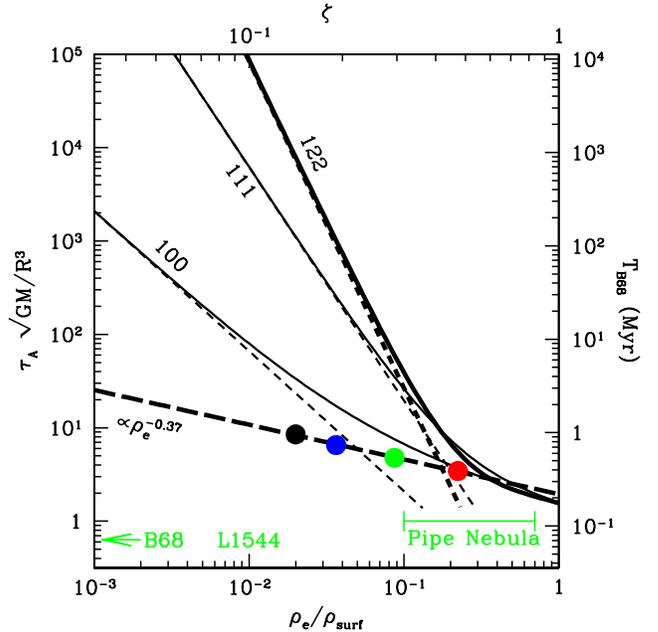}
}
\caption{The damping timescales in units of the dynamical time of the
  cloud, $(GM/R^3)^{-1/2}$, as a function of the ratio of the exterior
  and surface densities.  For reference the corresponding damping
  times in $\Myr$ is shown on the right-hand axis for a cloud similar
  to Barnard 68.  The solid lines show the damping
  timescale predicted by a linear mode analysis of the isothermal
  sphere for monopole (011), dipole (111) and quadrupole (122,thick)
  modes (see the appendices).  The asymptotic behavior of the linear
  mode analysis is shown by the short dashed lines for each, which may
  be obtained by treating the oscillations of the isothermal sphere as
  sound waves in a uniform density sphere (see the appendices).  The
  solid circles show the damping timescales measured via numerical
  simulations, and are colored to match the curves in Figures
  \ref{fig:ds} \& \ref{fig:sims}.  Finally, we show a power-law fit to the measured
  damping timescales, which have a power-law index of $-0.37$, and is
  considerably flatter than we might have expected (though these
  represent lower limits upon the decay timescale).  For comparison
  the density contrasts observed in some well known examples are also
  shown, including those for Barnard 68 ($10^{-5}$), L1544 ($10^{-2}$)
  and the cores in the Pipe Nebula ($0.1$--$0.7$).} \label{fig:tA}
\end{figure}

More directly applicable to the damping of oscillating gas spheres is
the case of the decay of a standing multipolar sound wave in a uniform
density sphere, discussed in section \ref{sec:SDT}.  This is distinct
from the 1D case in an important way: the oscillation wave vectors are
no longer only normal to the boundary.  As a consequence, the damping
rate is a strong function of not only $\zeta$ but also the multipole
structure of the underlying oscillation.  In particular, the
damping rate for the $l^{\rm th}$ multipole mode is proportional to
$\zeta^{2l+3}$ for small $\zeta$ (see, eqs.
\ref{eq:uniform_sphere_damping} \& \ref{eq:full_linear_damping}), which is a very strong function of
$\zeta$ for even low $l$!  This limiting behavior is also found in
semi-analytic calculations (section \ref{sec:FLP}) in which the linear
oscillation of an isothermal sphere is properly matched to outgoing
sound waves in the exterior.  The damping 
timescales (inverse of the damping rates) from both the uniform-sphere
approximation and the full linear mode analysis are shown in Figure
\ref{fig:tA} for the first few multipoles.  All of this implies that
the damping rates of oscillations of cores embedded in cold molecular
regions, where the density contrast between the core and the cloud is
near unity, will be much more rapid than those in cores isolated in hot,
lower-density regions.

\thispagestyle{empty}

\begin{figure}[t!]
\centerline{
\includegraphics[width=\columnwidth]{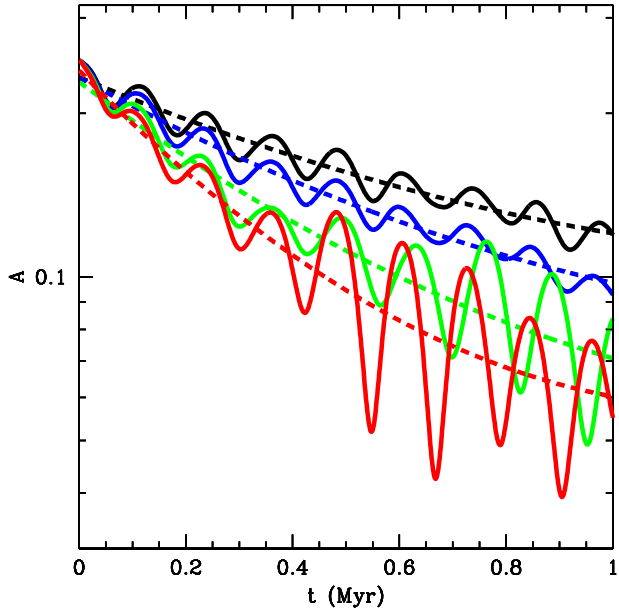}
}
\caption{The mode evolution for various external gas models.  The
  density contrast for each is roughly 50 (black), 25 (blue), 10
  (green) and 4 (red), and the lines are colored to match those in Figs.
  \ref{fig:ds} \& \ref{fig:tA}.  The decay time for the red curve is 0.43 Myr,
  compared to 1.05 Myr for the black curve.  Quadratic fits for each
  are shown by the dashed lines.  Matching these up to specific models
  of the mode decay is complicated by (i) the non-linear mode
  dynamics, and (ii) understanding the energy losses due to traveling
  waves in the external medium (see the appendices).} \label{fig:sims}
\end{figure}

However, this is not borne out by numerical simulations of high
amplitude oscillations.  This can readily be seen by the damping timescales
measured via the numerical simulations.  Figure \ref{fig:sims} shows
the evolution of the $n=1$, $l=2$, $m=2$ mode discussed in
\citet{Broderick2007}.  In all cases the initial amplitude was
0.25, and the evolution of the mode well fit by a decaying
exponential.  Oscillations on isothermal cores embedded in higher
density regions did indeed damp more rapidly, as is apparent in the
figure.  Explicit values of the decay timescale are presented in Table
\ref{tab:T}, though as in \citet{Broderick2007} these should be seen
as lower limits only due to coupling in the artificial atmosphere.
These are plotted as a function of the surface density contrast in
Figure \ref{fig:tA}, shown by the colored filled circles.  For
reference, we show the approximate surface density contrasts of
Barnard 68 \citep{AlvesLadaLada2001}, L1544
\citep{Ward-ThompsonMotteAndre1999} and typical for cores in the Pipe
Nebula \citep{Lada2008}. While there is a clear power-law dependence of
the damping timescale upon $\zeta$, with $\tau_A \propto
(\rho_e/\rho_i)^{-0.37}$ this dependence is considerably weaker than
that predicted by the linear analysis
($\tau_A\propto(\rho_e/\rho_i)^{-3.5}$).

This disparity may be understood in terms of the relative importance
of the excitation of sound waves in the exterior medium to the damping of
the oscillations.  In particular, it is notable that the numerically measured
damping timescales are always less, and for most surface density
contrasts considerably so, than those implied by the linearized
analysis of external sound wave excitation.  This is true even with
low amplitude oscillations (\eg, $A=10^{-4}$, for which the damping
timescale is roughly a factor of 3 larger than for $A=0.25$ ) and thus is likely due
to mode coupling in the transition region immediately outside the
cloud (seen in Fig. \ref{fig:ds} immediately outside $r/R=1$ as the
region of rapidly decreasing density).  Within this region the
densities are all within a single order of magnitude despite their
very different asymptotic densities at infinity (which differ by many
orders of magnitude).  As a consequence, the damping rates inferred
from the isolated isothermal spheres are only a weak function of the
properties of the external medium.  Thus, even for surface
density contrasts on the order of $1/4$, the decay timescale for the
quadrupolar oscillation is still larger than $4\times10^5\,\yr$,
corresponding to many oscillations and comparable to the inferred
lifetimes of globules.

\section{Conclusions} \label{sec:C}
Despite the expectations of the linear mode analysis of the embedded
isothermal sphere, the damping timescale of large-amplitude
oscillations of embedded Bonnor-Ebert spheres is only a weak function
of the density of the external medium.  This is a result of the
dominance of nonlinear mode--mode coupling in the damping of large
oscillations.  Even in cold molecular environments, the quadrupolar
oscillation discussed in \citet{Keto2006} and
\citet{Broderick2007} has a lifetime of at least $0.4\,\Myr$, and is
thus comparable to the inferred lifetimes of globules
\citep{Lada2008}.  This suggests that globules supporting
large-amplitude oscillations may be common even in these environments.

The presence of large-amplitude oscillations on starless cores has a
number of consequences for the physical interpretation of observations
of starless cores \citep{Keto2006,Broderick2007}.  Necessarily they
would imply that starless cores are stable objects, existing for many
sound-crossing times.  In addition, the signatures of collapse,
expansion and rotation in observations of self-absorbed molecular
lines (\eg, CS) are degenerate with molecular line profiles produced
by oscillating globules
\citep[see, \eg, figure 6 of~][]{Broderick2007}.
Consequently, it may be difficult to determine the true dynamical
nature of motions in a given globule or dense core from self-absorbed,
molecular-line profiles alone.  This suggests that studies of such
profiles in dense cores may be of only statistical value in
determining the general status of motions in dense core populations.

\acknowledgments
We would like to thank Mark Birkinshaw for bringing the problem of
damping starless core pulsations in dense media to our attention.

\appendix

\section{Damping of an Oscillating Cavity}
We will first review some basic facts about sound waves in uniform
media.  Beginning in subsection \ref{sec:1DT} we will discuss the
evolution of a one-dimensional standing sound wave as a result of the
excitation of waves in the external medium.  In subsection
\ref{sec:SDT} we discuss the application to oscillating uniform
density spheres, and treat the full linearized mode analysis of
isothermal spheres in section \ref{sec:FLP}.

\subsection{Equations of Motion}
The governing equations are the linearized continuity and the Euler
equations:
\begin{eqnarray}
\dot{\rho'} + \bmath{\nabla}\cdot \rho_0 \bmath{\delta v} &= 0\,,\nonumber\\
\dot{\bmath{\delta v}} + \frac{c_s^2}{\rho_0} \bmath{\nabla} \rho' &= 0\,,
\end{eqnarray}
where $\rho_0$ and $c_s$ are the unperturbed density and sound speed,
respectively, $\rho'$ is the Eulerian perturbation in the density and
$\bmath{\delta v}$ is the Lagrangian perturbation in the velocity
(which is identical to the Eulerian perturbation since the initial
velocity field is assumed to vanish).  These may be combined in the
normal way to produce the wave equation
\begin{equation}
\ddot{\rho'} - c_s^2 \nabla^2 \rho' = 0\,,
\label{eq:rho_wave_eq}
\end{equation}
where we have assumed that the background density is uniform.
Solutions will generally obey the dispersion relation $k^2 = c_s^2
\omega^2$, and may then be inserted into the Euler equation to obtain
the associated velocity perturbation.

\subsection{Decay of a One-Dimensional Sound Wave} \label{sec:1DT}
We will consider the decay of a sound wave traveling inside of a
uniform high density region (called the {\em interior} and denoted by
sub-script $i$'s) surrounded by a uniform low density region (called
the {\em exterior} and denoted by sub-script $e$'s).  The interior
wave oscillation is given by
\begin{equation}
\rho'_i = A_i \rho_i \sin(k_i x) \e^{-i\omega t}
\quad{\rm and}\quad
\delta v^x_i = - i A_i c_{s,i} \cos(k_i x) \e^{-i\omega t}\,.
\end{equation}
The exterior waves are outgoing, and we will focus upon the boundary
condition at the $+R$ boundary.  At this point the exterior sound wave is
given by
\begin{equation}
\rho'_e = A_e \rho_e \e^{i k_e x - i\omega t}
\quad{\rm and}\quad
\delta v^x_e = A_e c_{s,e} \e^{i k_e x - i\omega t}\,.
\end{equation}
At the interface we require (i) pressure equilibrium and (ii)
continuity of the displacement and hence velocity.  Since we have
assumed the $\rho_i$ and $\rho_e$ are constant, the first condition is
simply $\rho'_i c_{s,i}^2 = \rho'_e c_{s,e}^2$.  The second
is trivially $\delta v^x_i = \delta v^x_e$.  Thus
\begin{equation}
A_i \rho_i c_{s,i}^2 \sin(k_i R) \e^{-i\omega t}
=
A_e \rho_e \e^{i k_e R - i\omega t}
\quad{\rm and}\quad
- i A_i c_{s,i} \cos(k_i R) \e^{-i\omega t}
=
A_e c_{s,e} \e^{i k_e R - i\omega t}\,,
\end{equation}
give two complex equations from which we may determine $A_e$ and
$\omega$ as functions of $A_i$ and the background fluid quantities
(note that $k_i = \omega/c_{s,i}$ and $k_e = \omega/c_{s,e}$ by the
dispersion relations in each region).  In particular, note that by
taking the ratio of the two equations we find
\begin{equation}
\tan(k_i R)
=
- i \zeta\,,
\label{eq:1Dmatching}
\end{equation}
where $\zeta\equiv \rho_e c_{s,e}/\rho_i c_{s,i} = \sqrt{\rho_e/\rho_i}$ is the ratio of the
exterior and interior sonic impedances.  Generally this may be solved
to find that $k_i R = n\pi + (i/2) \ln\left[|1-\zeta|/(1+\zeta)\right]$, however we
will solve this explicitly in the $\zeta \ll 1$ limit to illustrate
how we will do this for spherical geometries later.

Let us begin by assuming that the damping rate is small.
Specifically, let us assume that $\omega_0\equiv\Re(\omega)$ is much
larger than $\gamma\equiv-\Im(\omega)$.  Then we may Taylor expand the
left-hand side of eq. (\ref{eq:1Dmatching}) around $\gamma=0$:
\begin{equation}
\tan(k_i R)
=
\tan\left(\frac{\omega_0 R}{c_{s,i}}\right)
-
\left.\frac{\partial\tan}{\partial \omega}\right|_{\omega=\omega_0} i
\gamma
+
\dots
=
0 - i\zeta\,,
\end{equation}
and thus, equating real and imaginary parts
\begin{equation}
\tan\left(\frac{\omega_0 R}{c_{s,i}}\right)
=
0
\quad\Rightarrow\quad\omega_0 = n \pi \frac{c_{s,i}}{R}
\end{equation}
and
\begin{equation}
\left[ 1 + \tan^2\left(\frac{\omega_0 R}{c_{s,i}}\right) \right] \frac{\gamma R}{c_{s,i}}
=
\zeta
\quad\Rightarrow\quad
\gamma = \frac{c_{s,i}}{R} \zeta\,.
\end{equation}
We may in principle then insert this into eq. (\ref{eq:1Dmatching}) to
obtain the relationship between $A_i$ and $A_e$.  However, we will
be concerned with only the damping rate here.

\subsection{Damping Timescale of an Oscillating Sphere} \label{sec:SDT}
In the previous section we discussed the simple problem of the
damping of a one-dimensional wave due to the excitation of outgoing
exterior sound waves.  In this section we will address the more
relevant problem of the damping timescale of multipolar oscillations
on a sphere.  While generally we would use the linearized analysis of
the oscillating isothermal sphere, here we will limit ourselves to
the spherical analog of the previous section: a sound wave in a
uniform density sphere surrounded by a uniform density exterior.

Before we can discuss the excitation of sound waves outside of an
oscillating sphere we must first determine the explicit form of these
waves.  We do this by separating the radial and angular dependencies
using spherical harmonics (primarily because these were used in
determining the mode spectrum of the isothermal sphere).  That is, we
let $\rho' = \e^{-i\omega t} \mathcal{R}(r) Y_{lm}(\hat{\bmath{r}})$, and
insert this into equation (\ref{eq:rho_wave_eq}), producing
\begin{equation}
\frac{1}{r} \frac{\partial^2}{\partial r^2} r \mathcal{R}
-
\frac{l(l+1)}{r^2} \mathcal{R}
+
k^2 \mathcal{R} = 0\,,
\end{equation}
where $k\equiv\omega/c_s$.
The general solutions of this equation are simply the spherical Bessel
functions:
\begin{equation}
\mathcal{R}(r) = a j_l\left(kr\right) + b n_l\left(kr\right)\,.
\end{equation}
However, we must still separate the inward and outward traveling
waves.  This naturally occurs if we consider spherical Bessel
functions of the 3rd kind, which are necessarily simply linear
combinations of spherical Bessel functions of the 1st and 2nd kind
($j_l$ and $n_l$, respectively):
\begin{equation}
h_l^{(1,2)}(z) \equiv j_l(z) \pm i n_l(z) \propto \frac{\e^{\pm iz}}{z^{l+1}}\,.
\end{equation}
Outwardly directed waves are given by $h_l^{(1)}(z)$ and inwardly
directed waves are given by $h_l^{(2)}(z)$.  In terms of this the
density perturbation of the exterior sound waves associated
with the $l,m$ multipole is
\begin{equation}
\rho'_e
=
A_e \rho_e \e^{-i\omega t} h_l^{(1)}(k_e r) Y_{lm}(\hat{\bmath{r}})
\,,
\end{equation}
where $A_e$ is a dimensionless wave amplitude.  For $l=0$ this
results in the standard spherical wave solutions, $\propto \e^{\pm ikr}/r$.
The velocity perturbation may be determined via the linearized Euler equation:
\begin{equation}
\delta v^r_e
=
- i \frac{c_s^2}{\omega \rho_e} \frac{\partial\rho'_e}{\partial r}
=
- i A_e c_{s,e} \e^{-i\omega t} \frac{\partial h_l^{(1)}}{\partial z_e}(z_e) Y_{lm}(\hat{\bmath{r}})\,,
\label{eq:sph_dv}
\end{equation}
where here and henceforth we have defined $z_e\equiv k_e r$ for convenience.  
In contrast, the interior wave must be regular at the origin, and is given by
\begin{equation}
\rho'_i
=
A_i \rho_i \e^{-i\omega t} j_l(k_i r) Y_{lm}(\hat{\bmath{r}})
\quad{\rm and}\quad
\delta v^r_i
=
- i A_i c_{s,i} \e^{-i\omega t} \frac{\partial j_l}{\partial z_i}(z_i) Y_{lm}(\hat{\bmath{r}})\,,
\end{equation}
where $z_i \equiv k_i r = z_e/\zeta$.

At the interface we again require pressure equilibrium and continuity
in the radial velocity:
\begin{equation}
A_i \rho_i c_{s,i}^2 \e^{-i\omega t} j_l(k_i r) Y_{lm}(\hat{\bmath{r}})
=
A_e \rho_e c_{s,e}^2 \e^{-i\omega t} h_l^{(1)}(k_e r) Y_{lm}(\hat{\bmath{r}})
\end{equation}
and
\begin{equation}
- i A_i c_{s,i} \e^{-i\omega t} \frac{\partial j_l}{\partial z_i}(z_i) Y_{lm}(\hat{\bmath{r}})\,.
=
- i c_{s,e} \e^{-i\omega t} \frac{\partial h_l^{(1)}}{\partial z_e}(z_e) Y_{lm}(\hat{\bmath{r}})
\end{equation}
As before, together with $\omega=k_i/c_{s,i}=k_e/c_{s,e}$ and
$z_e=\zeta z_i$, these are sufficient to determine $A_e$ and
$\omega$.  In particular, these imply
\begin{equation}
\frac{\partial \ln j_l}{\partial z_i} (z_i)
=
\frac{1}{\zeta} \frac{\partial \ln h^{(1)}_l}{\partial z_e} (z_e)
\end{equation}
The right-hand
side may be simplified in the $z_e \ll 1$ limit.  Since $z_i$ is
typically of order unity, this implies that $\zeta\ll 1$.  In this
regime, to leading order in small $z_e$,
\begin{equation}
\frac{1}{\zeta} \frac{\partial \ln h^{(1)}_l}{\partial z_e} (z_e)
\simeq
-\frac{l+1}{z}
+
i \frac{z_e^{2l}}{\left[(2l-1)!!\right]^2}\,,
\end{equation}
where $(2l-1)!! = (2l-1)\cdot(2l-3)\cdots 5 \cdot 3 \cdot 1$.

Again
let us begin with the ansatz that the damping rate is small for small
$\zeta$.  In which case
\begin{equation}
\left.\frac{\partial \ln j_l}{\partial z_i}\right|_{\omega_0}
-
\left.\frac{\partial^2 \ln j_l}{\partial z_i^2}\right|_{\omega_0} i\frac{\gamma R}{c_{s,i}}
+
\dots
=
-\frac{l+1}{\zeta^2 z_i}
+
i \frac{\zeta^{2l-1} z_i^{2l}}{\left[(2l-1)!!\right]^2}\,,
\end{equation}
and thus
\begin{eqnarray}
&&\displaystyle \left.\frac{\partial \ln j_l}{\partial z_i}\right|_{\omega_0}
=
\left[ \frac{j_{l-1}(z_i)}{j_{l}(z_i)} - \frac{l+1}{z_i} \right]_{\omega_0}
=
-\frac{l+1}{\zeta^2 z_i}\nonumber\\
&&\displaystyle \qquad\qquad\qquad\qquad\qquad\quad\Rightarrow\quad
\frac{j_{l-1}(z_i)}{j_{l}(z_i)} \simeq -\frac{l+1}{\zeta^2 z_i}
\quad{\rm and}\quad
\omega_0 \simeq \frac{c_{s,i}}{R} \mathcal{Z}_{nl}
\end{eqnarray}
where $\mathcal{Z}_{nl}$ is the $n^{\rm th}$ root of $j_l(z)$.  With
\begin{equation}
\left.\frac{\partial^2 \ln j_l}{\partial z_i^2}\right|_{\omega_0}
\simeq
-\left[\frac{j_{l-1}(z_i)}{j_{l}(z_i)}\right]^2
\simeq
- \frac{(l+1)^2}{\zeta^4 \mathcal{Z}_{nl}^2}\,,
\end{equation}
(where we used the fact that $\zeta\ll 1$) we find
\begin{equation}
\gamma
=
\frac{c_{s,i}}{R}
\frac{\zeta^{2l+3} \mathcal{Z}_{nl}^{2l+2}}{(l+1)^2 \left[(2l-1)!!\right]^2}\,.
\label{eq:uniform_sphere_damping}
\end{equation}

We note that this is quite different than the one-dimensional case.
In particular, the decay rate decreases rapidly with decreasing
$\zeta$, and is a strong function of $l$, with higher multipoles
decaying considerably more rapidly than lower multipoles.  This is a
consequence of our assumption that not just $\zeta \ll 1$, but $\zeta
z_i \ll 1$ (justified in our case), which implies that $\lambda_e \gg
R$.  That is, the exterior propagating sound wave necessarily knows
that the geometry is converging, with higher multipoles converging
more rapidly.

We should also emphasize that the scaling of the damping rate,
$\gamma$, with $\zeta$ is dependent primarily upon the structure of the
traveling sound waves in the exterior.  Rather, the properties of the
interior perturbation are encoded in the definition of the
$\mathcal{Z}_{nl}$, and thus the normalization. Therefore, the damping
rates associated with the linearized perturbations of the
pressure-supported isothermal sphere (Bonnor-Ebert sphere) should be
quite similar, differing only in the particular values of
$\mathcal{Z}_{nl} \simeq 2\pi R/\lambda_i$.  Indeed, in the next
section we find this to be the case.

\section{Damping of Oscillating Isothermal Spheres in the Linear
  Regime} \label{sec:FLP}
The rather surprisingly strong scaling of the damping timescale with
the density contrast motivates a more careful analysis of the damping
rates of a self-gravitating, pressure supported isothermal sphere.
The mode analysis of an isothermal sphere has been treated in
considerable detail elsewhere \citep[\eg,][]{Cox1980,Keto2006} and
thus we summarize it only briefly here.  The perturbed quantities are
most conveniently described by the Dziembowski variables
\citep{Dziembowski1971}
\begin{equation}
\bmath{\eta}(\bmath{r}) =
\left(\frac{\delta r}{r} ,
\frac{P'+\rho_0 \psi'}{\rho_0 g r} ,
\frac{\psi'}{g r} ,
\frac{1}{g}\frac{\partial \psi'}{\partial r}
\right)\,,
\end{equation}
which may be separated into radial and angular parts,
$\eta_i (\bmath{r}) = \eta_i (r) \e^{-i\omega t} Y_{lm}(\hat{\bmath{r}})$.
In terms of these, the linearized hydrodynamic equations are given by
\begin{mathletters}
\begin{eqnarray}
&&r \frac{\partial \eta_1}{\partial r}
=
(V-3) \eta_1 + \left[\frac{l(l+1)}{\sigma^2 C} - V\right] \eta_2 + V \eta_3
\label{eq:dz_a}\\
&&r \frac{\partial \eta_2}{\partial r}
=
\sigma^2 C \eta_1 + (1-U)\eta_2\\
&&r \frac{\partial \eta_3}{\partial r}
=
(1-U) \eta_3 + \eta_4\\
&&r \frac{\partial \eta_4}{\partial r}
=
U V \eta_2 + \left[ l(l+1) - UV \right] \eta_3 - U \eta_4\,,
\label{eq:dz_eqs}
\end{eqnarray}
\end{mathletters}
where
\begin{equation}
U \equiv \frac{\partial \ln r^2 g}{\partial \ln r}
\,,\quad
V \equiv - \frac{\partial \ln P_0}{\partial \ln r}
\,,\quad
C \equiv -\frac{G M}{r^2 g} \left(\frac{r}{R}\right)^3
\,,\quad
\sigma^2 \equiv \frac{R^3\omega^2}{G M}\,,
\end{equation}
are functions of the background equilibrium configuration.  These are
solved subject to a normalization condition $\eta_1(R)=1$ and the
inner boundary conditions:
\begin{equation}
\sigma^2 C \eta_1 \big|_{r=0} = l \eta_2 \big|_{r=0}
\qquad{\rm and}\qquad
\eta_4\big|_{r=0} = l \eta_3\big|_{r=0}\,,
\end{equation}
arising from regularity at the origin.  A third boundary condition is
a result from the continuity of the gravitational potential at the
sphere's surface,
\begin{equation}
\eta_4\big|_{r=R} = -(l+1) \eta_3\big|_{r=R}\,.
\end{equation}
If the exterior density vanishes, as was assumed in
\citet{Keto2006}, the final boundary condition is given by requiring
that the Lagrangian pressure perturbation,
\begin{equation}
\delta P
=
P' + \bmath{\delta r}\cdot\bmath{\nabla} P_0
=
\rho_0 g r \left( \eta_2 - \eta_3 - \eta_1 \right)
=
0\,.
\end{equation} 
However, we now properly match this onto the outgoing multipolar wave
solutions described in section \ref{sec:SDT}.  Namely,
\begin{equation}
\delta P
=
\rho'_e c_{s,e}^2 = A_e \rho_e h^{(1)}_l (z_e)\,,
\end{equation}
subject to
\begin{equation}
\delta r
=
\eta_1 R
=
\frac{c_{s,e}^2}{\omega^2 \rho_e} \frac{\partial\rho'_e}{\partial r}
=
A_e \frac{c_{s,e}}{\omega} \frac{\partial h^{(1)}_l}{\partial z_e}\,.
\end{equation}
Therefore,
\begin{equation}
\left. \eta_2 - \eta_3 - \eta_1\right|_{r=R}
=
\left.\eta_1 \frac{h^{(1)}}{\partial h^{(1)}/\partial z_e}
\frac{\rho_e c_{s,e} \omega}{\rho_0 g}\right|_{r=R}\,.
\end{equation}
As before we assume that the damping rate is small, set
$\omega=\omega_0 - i\gamma$ and employ the boundary condition
\begin{equation}
\left. \eta_2 - \eta_3 - \eta_1 \right|_{r=R}
=
\Re\left[ \eta_1 \frac{h^{(1)}}{\partial h^{(1)}/\partial z_e}
\frac{\rho_e c_{s,e} \omega}{\rho_0 g} \right]_{r=R, \omega=\omega_0}\,.
\end{equation}
The damping rate is then estimated by
\begin{equation}
\gamma 
=
-\Im\left[ \eta_1 \frac{h^{(1)}}{\partial h^{(1)}/\partial z_e}
\frac{\rho_e c_{s,e} \omega}{\rho_0 g} \right]_{r=R, \omega=\omega_0}
\bigg/
\left. \frac{\partial(\eta_2-\eta_3)}{\partial\omega} \right|_{r=R}\,,
\end{equation}
where we used the fact that $\partial\eta_1/\partial\omega = 0$ by the
normalization condition.  In practice
$\partial(\eta_2-\eta_3)/\partial\omega$ is evaluated by solving the
eigenvalue problem for the $\eta_i$ and $\sigma^2$ using
\begin{equation}
\left. \eta_2 - \eta_3 - \eta_1 \right|_{r=R}
=
\alpha \Re\left[ \eta_1 \frac{h^{(1)}}{\partial h^{(1)}/\partial z_e}
\frac{\rho_e c_{s,e} \omega}{\rho_0 g} \right]_{r=R, \omega=\omega_0}
\bigg/
\left. \frac{\partial(\eta_2-\eta_3)}{\partial\omega} \right|_{r=R}\,,
\end{equation}
and setting
\begin{equation}
\frac{\partial(\eta_2-\eta_3)}{\partial \omega}
=
\frac{\partial(\eta_2-\eta_3)/\partial\alpha}{\partial\omega/\partial\alpha}\,.
\end{equation}
This amounts to finding $\partial(\eta_2-\eta_3)/\partial\omega$
holding the other boundary conditions constant.

The results of this procedure are explicitly shown in Figure
\ref{fig:tA}.  Of particular note is that the intuition obtained from
the analysis of sound waves in a uniform density sphere is borne out
in the low $\rho_e/\rho_0(R)$ limit, for which
\begin{equation}
\gamma
\propto
\left(\frac{\rho_e}{\rho_0(R)}\right)^{(2l+3)/2}
=
\zeta^{2l+3}
\,.
\label{eq:full_linear_damping}
\end{equation}

\end{document}